\documentclass[journal]{IEEEtran} 
\usepackage{caption}
\usepackage{fmtcount}
\IEEEoverridecommandlockouts
\usepackage{url}
\usepackage{cite}
\usepackage{amsmath,amssymb,amsfonts}
\usepackage{algorithmic}
\usepackage{graphicx}
\usepackage{textcomp}
\usepackage{xcolor}
\def\BibTeX{{\rm B\kern-.05em{\sc i\kern-.025em b}\kern-.08em
    T\kern-.1667em\lower.7ex\hbox{E}\kern-.125emX}}

\usepackage{fancyhdr}
\renewcommand\thesection{\arabic{section}}

\renewcommand\thesubsection{\thesection.\arabic{subsection}}
\renewcommand\thesubsectiondis{\thesection.\arabic{subsection}}
\renewcommand{\thetable}{\arabic{table}}

\newcommand{\welive}{WE-LIVE~}

\begin{document}

\title{Spatio-temporal Latent Representations\\ for the Analysis of Acoustic Scenes \textit{in-the-wild}
}

\author{
    \IEEEauthorblockN{Claudia Montero-Ramírez$^1$}
    , \IEEEauthorblockN{Esther Rituerto-González$^{2, 3}$}
    , \IEEEauthorblockN{Carmen Peláez-Moreno$^{1, 4}$}
      \\
    \IEEEauthorblockA{$^1$\textit{Department of Signal Theory and Communications, Universidad Carlos III de Madrid, Spain}}
    \\
    \IEEEauthorblockA{$^2$\textit{Department of Psychiatry and Psychotherapy, University Hospital, Ludwig-Maximilian-University of Munich, Germany}}\\
    \IEEEauthorblockA{$^3$\textit{German Center for Mental Health (DZPG), partner site Munich-Augsburg, Germany}}\\
    \IEEEauthorblockA{$^4$\textit{University Institute on Gender Studies, Spain}}\\
      \IEEEauthorblockA{clmonter@pa.uc3m.es}, \IEEEauthorblockA{esther.rituertogonzalez@med.uni-muenchen.de},
    \IEEEauthorblockA{cpelaez@ing.uc3m.es}

}

\maketitle

\begin{abstract}
In the field of acoustic scene analysis, this paper presents a novel approach to find spatio-temporal latent representations from in-the-wild audio data. By using \welive, an in-house collected dataset that includes audio recordings in diverse real-world environments together with sparse GPS coordinates, self-annotated emotional and situational labels, we tackle the challenging task of associating each audio segment with its corresponding location as a pretext task, with the final aim of acoustically detecting violent (anomalous) contexts, left as further work.
By generating acoustic embeddings and using the self-supervised learning paradigm, we aim to use the model-generated latent space to acoustically characterize the spatio-temporal context.
We use YAMNet, an acoustic events’ classifier trained in AudioSet to temporally locate and identify acoustic events in \welive. 
In order to transform the discrete acoustic events into embeddings, we compare the information-retrieval-based TF-IDF algorithm and Node2Vec as an analogy to Natural Language Processing techniques. 
A VAE is then trained to provide a further adapted 
latent space. The analysis was carried out by measuring the cosine distance 
and visualizing data distribution via t-Distributed Stochastic Neighbor Embedding, revealing distinct 
acoustic scenes. Specifically, we discern variations between indoor and subway environments. Notably, these distinctions emerge within the latent space of the VAE, a stark contrast to the random distribution of data points before encoding.
In summary, our research contributes a pioneering approach for extracting spatio-temporal latent representations from in-the-wild audio data.

\end{abstract}


\section{Introduction}

An \textit{acoustic scene} refers to the environment or ambience in which different sound sources are produced and perceived. Focusing on the perception of these sounds, we could say that it is an \textit{affective acoustic scene} since it elicits emotions in humans. This elicitation is caused by previous experiences evoked by such acoustic scene and therefore, emotions can be detected as they are linked to an acoustic scene. Previous works based on affective computing have shown that by characterising these emotions in the acoustic scene, it is possible to detect risk situations, such as in the context of gender-based violence (GBV) \cite{Rituerto}. As the acoustic context is able to elicit emotions \cite{emociones}, characterizing it is the first step towards predicting risk situations in a real environment. So, as a first step, in this paper we want to contribute to the acoustic scene analysis field by characterizing acoustic spatio-temporal environments in-the-wild as a preliminary task, in order to learn useful representations following the self-supervised learning paradigm. 

As previously mentioned, context is crucial in emotion elicitation, of which the spatio-temporal acoustic scene is part. 
In similar studies such as "What Makes Paris Look like Paris?" \cite{Paris} where the context is analyzed from the visual perspective, in this study the aim is to analyze it from the acoustic point of view. In any real-life situation, the acoustic environment significantly affects emotion induction, e.g., in situations of extreme traffic or loud noise, the sound of the waves on a paradisaical beach or even passive background music. In fact, over the years, multiple audio-visual materials that are focused on the induction of emotions have been developed, such as film soundtracks \cite{soundtracks}, \cite{emdb}. Nevertheless, the induction of emotions through the acoustic environment in real situations has not been developed in such a far-reaching way, hence the aim to characterize the acoustic environment in real space-time using an in-the-wild dataset that also has emotion labels.

Our main goal with this research paper relies on the characterization of unlabelled in-the-wild acoustic scenes. In order to characterize in-the-wild acoustic scenes we take into account the several acoustic events or sounds that occur in each of them. To this end, we compare two pre-trained acoustic event detection models trained in large datasets: YAMNet \cite{YAMNet} and PANNs CNN14 model \cite{PANNs}. Following with the acoustic scene characterization, in particular, we use methods based on Natural Language Processing (NLP) and information retrieval, such as the TF-IDF (term frequency-inverse document frequency) algorithm \cite{TFIDF} --for the sake of balancing the occurrence of sounds in contexts-- and the Node2Vec algorithm \cite{Node2Vec}, as an analogy to Word2Vec \cite{Word2Vec} --in order to capture semantic relationships between acoustic sounds--, both of them with the final aim of obtaining a vector representation for each acoustic context. 
Since the emotional annotations in \welive are very sparse, we obtain the previous vector representations, following the self-supervised learning paradigm to utilize original unlabeled data to create additional context information \cite{VAE_selfsup}. 
In particular, we use a Variational AutoEncoder (VAE), to discriminate different contexts in an explicit latent space. This latent space is continuous, which means that it can learn to represent different characteristics of the input embeddings in different parts of the space \cite{VAE_latent}.

To carry out this study we use the UC3M4Safety \welive dataset that is the successor in-the-wild of the WEMAC dataset \cite{WEMAC}. This new dataset contains approximately one week of almost-continuous recordings of 14 users. Within it, audio, physiological signals, geolocation signals (GPS) and self-annotated spatio-temporal locations and emotions are captured. For this paper, specifically, the audio signals and associated GPS cell location tags are utilized. Additionally, self-reported situational labels are employed as supplementary information about GPS cell location tags. By using the cosine-similarity function 
and t-Distributed Stochastic Neighbor Embedding (t-SNE) dimensionality reduction algorithm representations of the VAE's latent space, we find that the self-supervised learning-based model's latent space shows the similarities and disparities between different acoustic scenes as expressed in the dataset. It is important to note that the location labels are non-binding since 1) the dataset was recorded in-the-wild and the user's spatial itineraries are unscripted and thus unpredictable, and 2) the battery consumption savings algorithm deployed in the recording devices prevents a continuous GPS signal transmission. Thus, the less-than-perfect quality of the location tags.

The remainder of this paper is organized as follows: Section \ref{RW} provides the state-of-the-art in the fields of acoustic event classification and, to a minor extent, the relationship of the acoustic context to emotion induction. As currently there are no available publications describing \welive, Section \ref{DAT} provides useful information about it. Section \ref{MET} describes the steps that were followed to carry out the experiments, and Section \ref{RE} provides the results, which are discussed in Section \ref{DIS}. Finally, Section \ref{CON} lists several conclusions with future work being outlined in Section \ref{FW}.

\section{Related work} \label{RW}

In the research field of acoustic scene classification, a fast increase of scientific publications has been observed in the last decade due to progress in the area of deep learning \cite{Jakob}. Among these developments, we find the DCASE community\footnote{https://dcase.community/challenge2021/task-acoustic-scene-classification}, which has also launched an acoustic scene classification challenge in 2023. Looking at recent works, \emph{Kushwaha, S. et al.} \cite{Kushwaha} have published in August 2023 a paper that aims to classify acoustic scenes from an embedding space obtained through an encoder, using ensembled acoustic event labels and unlabelled audio signals, in a very similar way to what we aim to do in this paper. Other papers published in previous years, such as \cite{acoustic_scene} by authors \emph{Barchiesi, D. et al.}, have already shown the possibilities of characterizing acoustic scenes based on different feature extraction methods, statistical models, meta-algorithms, etc. One such method is the classification of acoustic scenes based on acoustic sounds or events, a study that was developed in 2010 \cite{events} by \emph{Heittola, T. et al.}, whose premise was to classify audio contexts--classification between everyday environments--, according to the occurrence of acoustic events. 

Similarly, numerous Natural Language Processing (NLP) techniques were developed in recent years based on countings. 
These techniques include the TF-IDF algorithm \cite{TFIDF}, which has been used as an analogy between NLP and affective computing in several studies \cite{Clara} \cite{nlp_affective}. Specifically, a recent study \cite{Clara} uses the TF-IDF algorithm to obtain a vector of emotional representations of acoustic scenes based on acoustic events, so that sounds are treated as words and audio signals as documents, establishing an analogy with NLP and information retrieval. 

In addition, a large-scale dataset of manually annotated audio events has been developed: AudioSet \cite{AudioSet}, and related to that, multiple pre-trained models have been developed for the classification of acoustic events, taking as labels the AudioSet ontology, such as YAMNet \cite{YAMNet}, VGGish \cite{VGGish} or PANNs \cite{PANNs}. These models are candidates to offer a promising alternative for the characterisation of the acoustic environment, in order to be transferred to other domains and tasks, as it has already been shown in the field of affective acoustic scene analysis \cite{Clara}.

As there are multiple datasets focused on emotion elicitation \cite{Deng} but few of them contain a sufficiently numerous representation of \textit{fear} \cite{Celia} our team, UC3M4Safety\footnote{https://www.uc3m.es/institute-gender-studies/UC3M4Safety}, recorded a dataset (WEMAC) for fear recognition in the EMPATIA-CM\footnote{https://www.uc3m.es/instituto-estudios-genero/EMPATIA} project, which is freely available \cite{WEMAC}. As a continuation, 
another dataset called \welive has been developed. This new dataset contains data from $14$ users recorded during approximately one week in-the-wild, collecting their daily routines. It includes audio, physiological signals, GPS geolocation and a set of labels related to the users' emotions and location. 

Another contribution of the project to previous work is the addition of other NLP techniques for the representation of acoustic events. The rise of these models in recent times has led to their application in audio-related affective computing domains \cite{nlp_affective}. Specifically, in the previous work related to the acoustic context of the UC3M4Safety team, the TF-IDF algorithm was used \cite{Clara}.

However, looking back at the DCASE community and the classification of acoustic scenes, since the EMPATIA-CM project aims at predicting risk situations linked to GBV in-the-wild, there are two main differences in the objectives of the two projects. The first of them is the importance of characterizing indoor places in our project, especially 'home', as in the GBV it is a scenario where risk situations can often happen \cite{gbv_home}. The second and most important difference is that in this project the aim is not to classify acoustic environments, but to characterize them in a latent space in a self-supervised way. As the objective is to perform inference in-the-wild in real-time, our model has to be able to adapt to changing acoustic environments as they emerge, as well as to find differences beyond spatial location, including context.

\section{\welive Dataset} \label{DAT}

The predecessor of \welive is WEMAC, a laboratory recorede database aimed at detecting realistic emotions from a multimodal point of view in women \cite{tesis_esther}. However, as it is far away from real-life conditions, the UC3M4Safety team created the “Women and Emotion in real LIfe affectiVE computing dataset”, \welive\cite{tesis_esther}. 

The primary aim of \welive is to collect physiological, physical, and contextual data from women in an authentic and uncontrolled environment \cite{tesis_esther}. Additionally, it involves labeling their emotional responses to daily life events. Users can also tag their current location: home, work, transport, etc. This provides additional descriptive categorical location labels that are used in this study \cite{tesis_esther}. This comprehensive data collection is facilitated through the utilization of the Bindi system, comprising a wristband, pendant, mobile application, and server \cite{Bindi}. Via a Bluetooth® link to the mobile phone, the information captured by Bindi is transmitted to a secure and encrypted server \cite{tesis_esther}.

The dataset is composed of multimodal data from 14 women volunteers including Gender-based Violence Victims (GBVV), captured almost-continuously for $7-10$ days. All participants were enlisted via social media promotions and by reaching out to students and researchers at the university \cite{tesis_esther}. Volunteers were also requested to complete a standard questionnaire to document their everyday activities and daily routine, which are subsequently categorized by a specialized psychologist based on their relevance to the study. Throughout the experiment, participants have the option to discontinue their participation at any time \cite{tesis_esther}.

\section{Methodology} \label{MET}

\subsection{Location Data}

In this research, data from the $14$ female volunteers obtained from the \welive dataset were used. After a first inspection, the data from one of them was discarded due to insufficient audio data for the experimentation.

In order to discretely tag, label or categorize the locations in our data, we analyzed the spatially continuous GPS data in \welive Database. We used a hexagonal grid over the terrestrial map in order to delimit the GPS coordinates into more discrete locations. The edge of each hexagon was set to $0.0015$ in terms of the magnitude between latitude and longitude, which results in approximately $60$ meters in radius, as a common parameter for all users. Besides, in order to have a discrete number of locations for each user, we selected only the $10$ places --hexagons-- where the user had spent the most minutes recorded by the GPS signal. Each audio recorded within a hexagon was interpreted as the categorical location that the user had noted in that hexagon (note, there were unlabeled hexagons). In the dataset, there are $34,6 \pm 16,5$ hours of unlabeled audio data recorded per user, while each user has $2,6 \pm 1,5$ hours of annotated audio related to the GPS cell location tags.

\begin{figure}[htbp]
    \centerline{\includegraphics[width=0.5\textwidth]{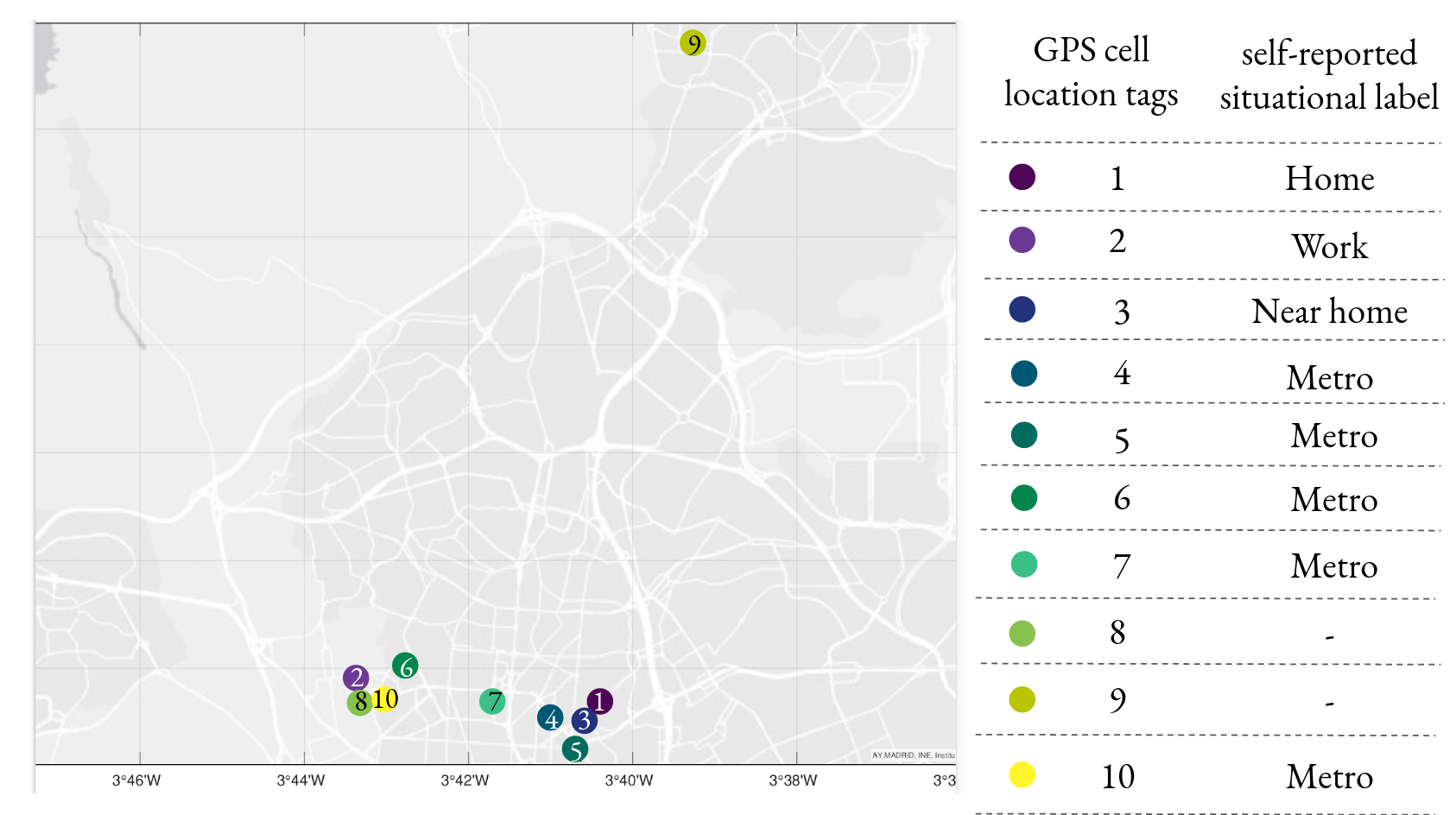}}
    \caption{Map of top 10 most frequented cells by a volunteer. In the left column, the corresponding index of each of the cells based on the GPS cell location tags. In the right column, the corresponding additional information based on the self-reported situational label.}
    \label{maps}
\end{figure}

Due to both the limited quality of the GPS data and the limited number of location-tagged data, for the visualization of results and conclusions, the self-reported situational labels by the volunteers were used as supplementary information to the GPS cell location tags. This can be seen in Fig. \ref{maps}, where the colored circles indicate the centroids of the GPS cell location tags. These colored circles have been enlarged for privacy preservation purposes.

\subsection{Sound Event Detection} \label{SED}

To categorize acoustic scenes based on their spatio-temporal location, we first perform a discrete acoustic event detection within them. To this aim, two pre-trained models were considered: YAMNet and PANNs CNN14.

Both YAMNet and PANNs CNN14 are pre-trained acoustic event detection models trained in a large-scale dataset that output multi-label soft probabilities of appearance for each of the AudioSet categories of sounds. 

In this case we have stuck to the computation time measured for 100 random audios from our dataset. Finally, YAMNet was chosen as the acoustic event detection model since its inference time was notably lower given the limited hardware constraints in this experiment.

The audio recordings of \welive were cut to one minute with the intention of being long enough to capture enough acoustic events to describe a scene, and short enough for the volunteers to remain in the same place for that time. For the signal pre-processing, the audio signal is normalized and converted to 16 kHz mono. A patch hop of $1$ second and a patch window of $1$ second were chosen as windowing parameters, resulting in a matrix of $60x521$ for each audio. The first dimension, $60$, refers to the the duration of the audio in seconds, and $521$ is the number of acoustic events or classes that YAMNet is able to distinguish. In that way, YAMNet is outputting the probability of each of the $521$ acoustic events appearing for each second. 

Finally, since the next step is to use analogies with information retrieval and NLP, acoustic events' probabilities are converted into binary values and then treated as if they were words. To this end, for each of the users the $99^{th}$ percentile of all acoustic events probabilities was calculated and triggered as the activation threshold, so that for all users approximately $5$ acoustic events were triggered per second. 

\subsection{Embeddings: an analogy with Information Retrieval and Natural Language Processing} \label{EMB}

In order to obtain a vectorial representation of
the acoustic scenes based on an analogy with techniques that are often used in information retrieval and NLP, two algorithms were employed: TF-IDF and Node2Vec.

TF-IDF\cite{TFIDF} is a statistical method that can quantify the importance or relevance of string representations in a document amongst a collection of documents\cite{TFIDF2}. In this case, the measure would be the relevance of an acoustic event in an audio with respect to all audios. The algorithm has two components: the term-frequency (TF) and the inverse document frequency (IDF) \cite{TFIDF3}. The term-frequency measures the frequency of a particular acoustic event, while the document frequency looks at how common an acoustic event is by counting in how many audios it appears and is used inversely. 

However, as TF-IDF is a count-based method, it is losing information about the semantic relationship of these acoustic events. To this end, the AudioSet ontology itself contains information that could be useful, as it captures this type of semantic relationship. The AudioSet ontology  consists of $615$ nodes corresponding to sounds and $482$ edges that relate them semantically \cite{AudioSet}. For example, there is a parent node called 'Music' from which different types of acoustic events related to music emerge. For this paper, besides filtering or grouping these nodes, a predictive algorithm (similar to Word2Vec\cite{Word2Vec}) is used, but instead of using the corpus, the relations of the AudioSet ontology itself are used. First of all, a relational tree is built containing the information of the ontology, and then the predictive algorithm Node2Vec\cite{Node2Vec} is applied. A $5x521$ matrix is obtained related to Node2Vec embeddings, so that for each acoustic event there is a 5-dimensional vector representation.

\begin{figure}[htbp]
    \centerline{\includegraphics[width=0.45\textwidth]{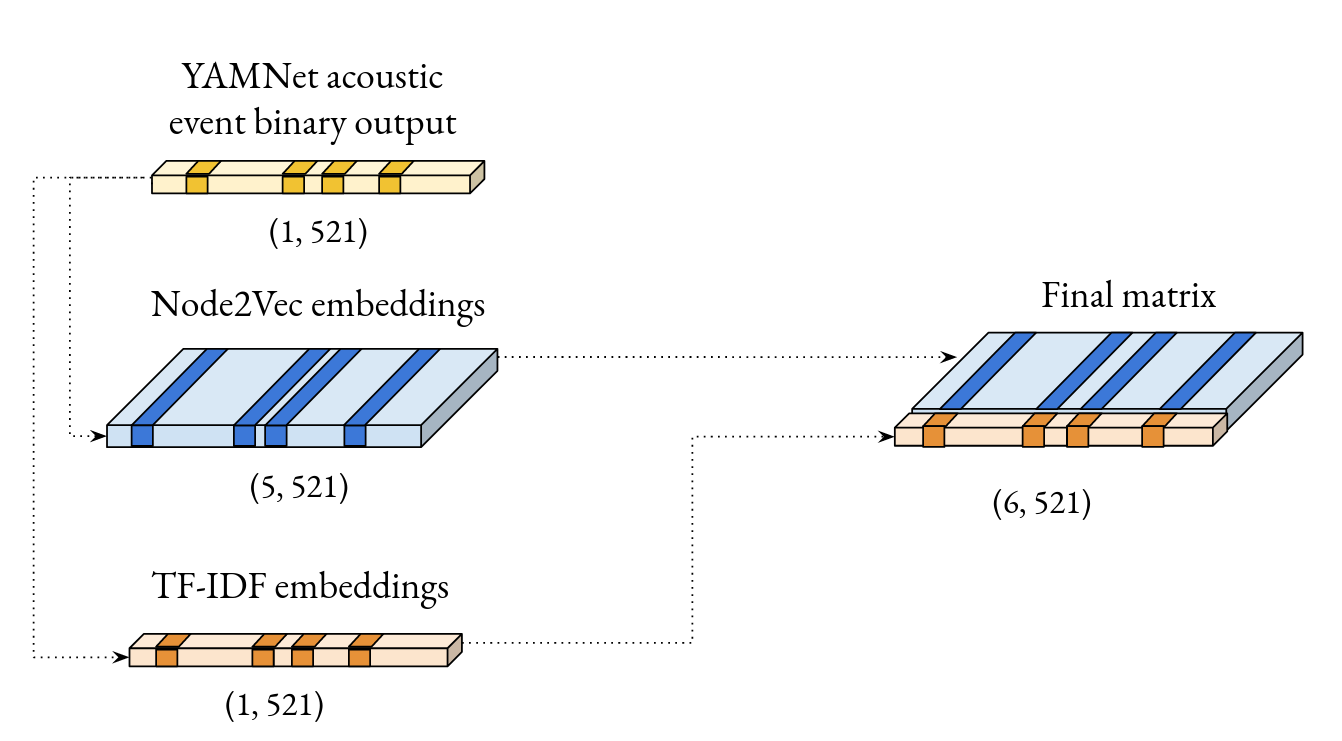}}
    \caption{Final embedding matrix representation for each one minute audio. Dark colours represent triggered acoustic events' vectorial representation, non-zero values. For TF-IDF and Node2Vec embeddings, each of them has its respective non-zero values, whereas for the YAMNet acoustic event binary output, the values are 1. Light colours represent non-triggered acoustic events, whose values are 0.}
    \label{embeddings}
\end{figure}

At the end, for each audio TF-IDF and Node2Vec embeddings are concatenated and a $6x521$ matrix is obtained, where the first dimension belongs to the TF-IDF representation and the remaining $5$ dimensions are related to the Node2Vec embeddings, as shown in Fig. \ref{embeddings}. 

\subsection{Variational AutoEncoder's Latent Acoustic Representation} \label{VAE_methodology}

Using this type of embeddings results in large and especially empty matrices. When a VAE is dealing with these sparse matrices, it may have inferior performance due to insufficient capability to model overdispersion and misspecification, among other risks \cite{VAE1}. In order to mitigate these possible issues two choices were made. 

The first one is the use of momentum in the gradient descent algorithm, as this speeds up training and mitigates the saturation in a local minimum. Besides that, the effectiveness of other traditional gradient descent methods is reduced when the input data is sparse\cite{VAE_alex}. Therefore, the stochastic gradient descend with momentum was employed, as shown in Eq. (\ref{momentum})\cite{pytorch}.

\begin{equation}
\begin{aligned}
\label{momentum}
v_t &= \beta \cdot v_{t-1} + \nabla J(\theta) \\
\theta_t &= \theta_{t-1} - \alpha_t \cdot v_t
\end{aligned}
\end{equation}

where $\theta$ are the parameters, $\nabla J(\theta)$ is the gradient of the cost function of the VAE (given by Eq. (\ref{vae_loss})) with respect to the parameters, assuming Gaussian distributions \cite{selfsupervised}. $v$ is the velocity, $\alpha$ is the learning rate and $\beta$ is the momentum, that in this case was set on $\beta=0.9$.

\begin{equation}
\begin{aligned}
\label{vae_loss}
\mathcal{L}_{VAE} = -\mathbb{E}_{q(z|x)}\left[ -\log(p(x|z)) + \text{KL}(q(z|x) | p(z)) \right]
\end{aligned}
\end{equation}

The second choice for dealing with sparse matrices was to select an adaptive learning rate, since in this way, the training converges faster \cite{VAE_alex}. The formulae used to obtain a decreasing and adaptive Learning Rate (LR) is shown in Eq. (\ref{exponentialLR}), which relates to exponential LR \cite{pytorch2}.

\begin{equation}
\begin{aligned}
\label{exponentialLR}
\alpha_t = \alpha_0  \cdot \gamma^{t}
\end{aligned}
\end{equation}

where $\gamma=0.99$ and $\alpha_0=1 \times 10^{-5}$.

\begin{figure}[htbp]
    \centerline{\includegraphics[width=0.455\textwidth]{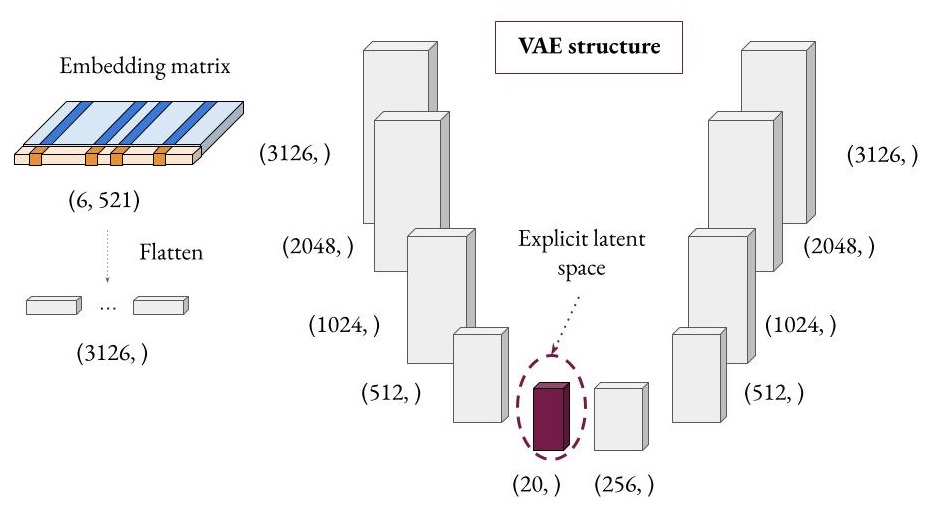}}
    \caption{Linear VAE structure.}
    \label{VAE_structure}
\end{figure}

As there is no spatial relationship between the resulting embeddings in Section \ref{EMB}, they were flattened into a 1-dimensional vector and a linear VAE \cite{LVAE} was applied, whose structure is shown in Fig \ref{VAE_structure}. The ReLu activation function was used in all layers, except for the last layer of the decoder, where the hyperbolic tangent (tanh) was employed. It has also been used in order to visualize the latent space, as will be shown in Section \ref{RE}, but not for the training or the inference. In addition, batch normalization has been applied between all layers. 

One VAE has been trained for each one of the users and 15\% randomly selected each user's audios were used for testing, while the remaining 85\% were employed for training. The number of training epochs was adjusted for each user depending on the amount of user data, using early stopping criteria. The following considerations were applied for each user: 
\begin{itemize}
    \item A comparison of the cosine distance (Eq. (\ref{cosine_distance}), \cite{Clara}) related to the GPS cell location tags, between the input embeddings (Section \ref{EMB}) and the latent space of the VAE (Section \ref{VAE_methodology}). A cosine distance matrix was computed to compare the input embeddings with each other, as well as another matrix for comparing the vectors within the latent space.

    
    \item In order to visualize in some way the latent space data related to pseudo-labeled locations, the t-SNE dimensionality reduction algorithm was applied \cite{TSNE_original} \cite{TSNE}. It was performed after applying a hyperbolic tangent activation function to the latent space data, for all training epochs. It was also used when making an inference with the test data, before and after training. 

    \item Loss function (Eq. (\ref{vae_loss})) evolution during epochs. 

\end{itemize}

\begin{equation}
\begin{aligned}
\label{cosine_distance}
\text{similarity}(x, y) = 1 - \cos(\theta) = 1 - \frac{x \cdot y}{\|x\| \|y\|}
\end{aligned}
\end{equation}

\section{Results}\label{RE}

Results related to the comparison of the inference time between YAMNet and PANNs CNN14 (Section \ref{SED}) are shown in TABLE \ref{comparision_time}, where the inference time required by YAMNet is lower.

\begin{table}[h]
\captionsetup{labelsep=period, font=small, labelfont=bf}
\begin{tabular}{|c|c|}
\hline
\textbf{Model} & \textbf{Preprocessing and Inference Time (seconds)} \\    & \textbf{mean $\pm$ std}\\ 

\hline
YAMNet & 0.66 $\pm$ 0.19 \\
\hline
PANNs CNN14 & 10.73 $\pm$ 0.65\\
\hline
\end{tabular}
\centering
\centerline{}
\caption{Comparision between YAMNet and PANNs CNN14 preprocessing and inference time in seconds, expressed as mean $\pm$ standard deviation, for 100 random audios of the \welive dataset.}
\label{comparision_time}
\end{table}

\begin{figure}[htbp]
    \centerline{\includegraphics[width=0.45\textwidth]{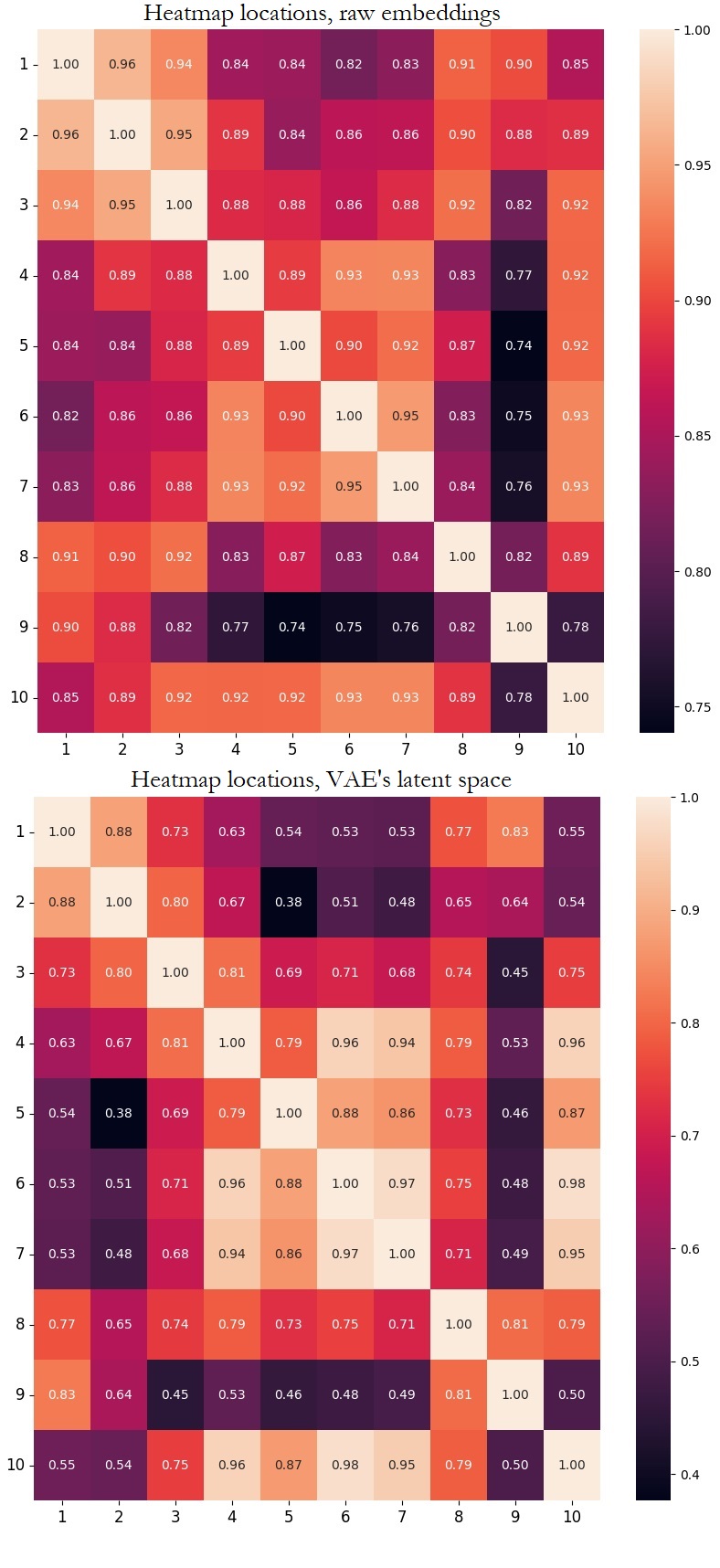}}
    \caption{Cosine distance between pseudo-labeled vector representations, for one user. Above, raw embeddings (Section \ref{EMB}) as vector representations. Below, VAE's latent space (Section \ref{VAE_methodology}) as vector representations. Matching the locations with those labeled by the user: 1-Home, 2-Work, 3-Near home, 4,5,6,7 and 10-Metro, 8 and 9-Unknown.}
    \label{heatmaps}
\end{figure}

Cosine distance between pseudo-labeled vector representations is shown by two colour heatmaps for one user in Fig. \ref{heatmaps}, where the above one is related to raw embeddings as vector representations (Section \ref{EMB}) --which is a concatenation of the TF-IDF and the Node2Vec embeddings-- and the below one uses VAE’s latent space (Section \ref{VAE_methodology}) as vector representations. As shown in Fig. \ref{heatmaps} vector representations in the explicit latent space show a higher discriminative ability than raw embeddings, showing a larger range of similarity metric values (Eq. \ref{cosine_distance}) and also two clusters are found: one related to indoor locations (home = 1 and work = 2) and the other referring to metro locations (4, 5, 6, 7 and 10), locations 8 and 9 are unknown, and location 3 is somewhere near home. 

\begin{figure}[htbp]
    \centerline{\includegraphics[width=0.5\textwidth]{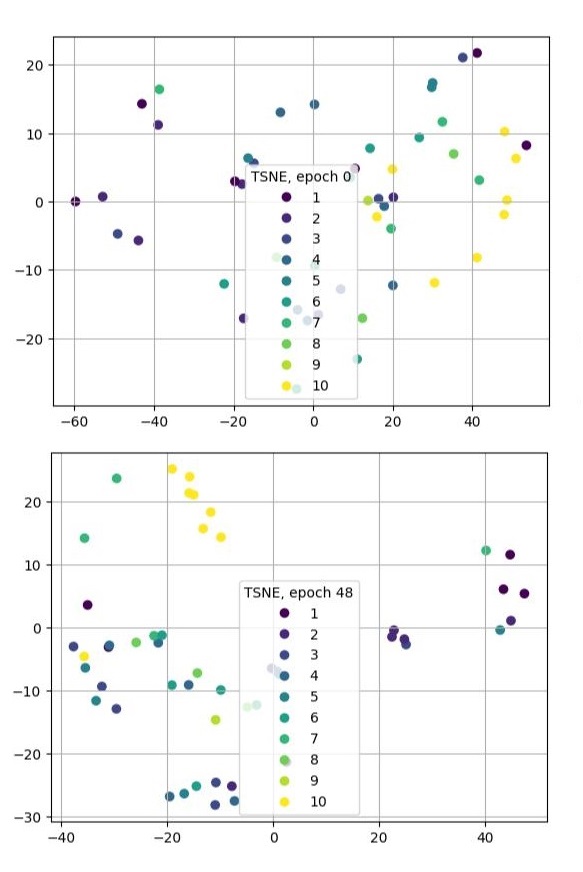}}
    \caption{t-SNE algorithm visualization during training for one user. Above, the distribution for epoch 0. Below, for epoch 48.}
    \label{scatter}
\end{figure}

Latent space evolution for training data during epochs by applying the t-SNE algorithm is shown in Fig. \ref{scatter} for one user, by two scatters in two different epochs. In the first epoch, a random spatial distribution of data is observed, while as the epochs increase, the data are regrouped. E.g., a clustering of data related to location number 10 is observed. It should be noted that the t-SNE algorithm uses random initialisation, so these plots are not conclusive \cite{TSNE_random}. 

\begin{figure}[htbp]
    \centerline{\includegraphics[width=0.45\textwidth]{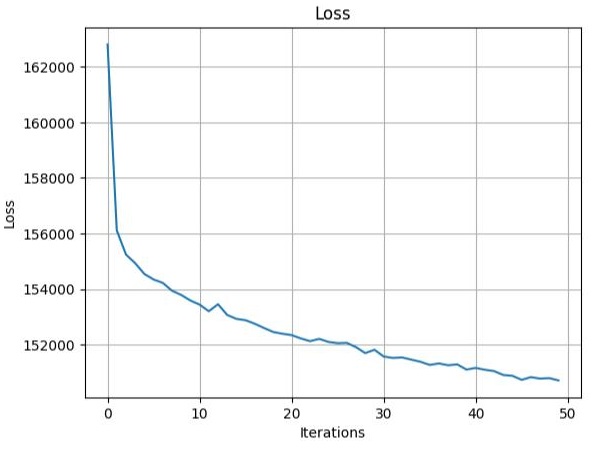}}
    \caption{VAE loss function evolution during epochs example for one user.}
    \label{loss_graph}
\end{figure}

The evolution of the loss function is shown as an example for one user in Fig. \ref{loss_graph}. It can be shown how loss function decreases as the training epochs progress.

\section{Discussion} \label{DIS}

For the discussion section, the four main contributions of this project are analyzed: the performance of the in-the-wild experiment and its challenges, the comparison between different pre-trained models of acoustic event detection, the implementation of other NLP techniques for obtaining acoustic vector representations and the addition of a VAE in order to obtain a self-supervised discriminative latent space.

\subsection{Dealing with the complexities of conducting in-the-wild experiments} \label{ITW}

In contrast to previous work that tackled the characterisation of acoustic scenes for emotion induction using NLP and information retrieval techniques \cite{Clara}, in this instance, we have an in-the-wild experiment. Working with audio signals captured in real time and unforseen places can lead to complications, such as movement artefacts, microphone saturation or the influence of device placement (e.g. in the amplitude of the audio signal).

When running these in-the-wild audio samples through the pre-trained YAMNet network, it could systematically misclassify certain sounds, resulting in a bias in the outcomes. This should be a minor problem when using the TF-IDF algorithm, as it relies uniquely on the corpus to obtain the representations. However, Node2Vec uses the AudioSet ontology itself to build the vector representations, so that it does not overlook the bias. 

This problem could be solved using the Word2Vec algorithm since it would capture the proximity relationships between acoustic events based exclusively on the corpus. Nonetheless, using this approach, the main purpose for which ontology-based Node2Vec was applied, which is to capture the semantic relationships, is lost. For example, one of the expected outcomes is to group all the descendants of the music node in a close vector space, a task that would not be possible by using Word2Vec. 

Thus, in order to combine the TF-IDF and Node2Vec representations, a very high threshold is needed, i.e., the aim is to reduce the bias as much as possible, so that YAMNet is sure that a certain sound has been produced with a high probability. For that reason, the $99^{th}$ percentile of all acoustic event probabilities is used as the trigger threshold. 

Setting audio aside, another challenge of conducting this in-the-wild experiment is the uncertainty regarding the spatiotemporal context for each audio sample. GPS is a highly battery-consuming device, so it is set off for much longer than the audio recording, resulting in few audio samples reliably related to specific locations. In addition, GPS geolocation is not always accurate, resulting in many false tags. For example, when a user goes into an underground station, the GPS stays for several minutes recording at that location, whereas the user is actually moving. 

Hence, the term 'pseudo-label' is attributed to the locations, due to the presence of numerous audios with diverse content within those locations. Furthermore, owing to their diminished quality, it is essential to underscore that these labels cannot be employed for a supervised learning model. Instead, an unsupervised approach must be pursued, with these labels serving merely as guiding indicators for result visualization. Moreover, any outcome linked to an accuracy or any metric using these labels would not be significant.

\subsection{Comparison of pre-trained models for acoustic event detection} \label{COMP}

Furthermore, a comparison between two pre-trained models with large dataset for acoustic event detection has been carried out in Section \ref{SED}. The evaluation was based on the execution times for this particular dataset, as shown in TABLE \ref{comparision_time}.
Although other parameters could have been analyzed, such as the output agreement of both networks, for the future purpose of this study (detecting risk situations in real time) computation time is one of the crucial parameters. 

However, there are many pre-trained models for this task, including the already cited VGGish \cite{VGGish}, the DCASE community audio event detection challenges and other methods compared in PANNs paper \cite{PANNs}, such as CNN6 or CNN10, whose performance could also be evaluated for this task.

\subsection{Analogy with Information Retrieval and NLP: Capturing semantic relationships using Word2Vec}

One of the contributions of this paper to previous similar works \cite{Clara} is the addition of other NLP methods in order to achieve vector representations. In this case, besides using the count-based method TF-IDF \cite{Clara} a prediction-based method was also added in order to capture semantic relationships between AudioSet nodes, as it is Node2Vec. Whereas the choice of Node2Vec is largely explained in Section \ref{ITW}, it should be added that this method is implemented as an alternative to node clustering or filtering. 

In previous work, e.g. \cite{Clara}, a clustering of all descendant nodes of 'Music' was performed. In this case it is not necessary to make such grouping, since in vector representations the neighbourhood of these nodes will be reflected. For this specific task Node2Vec representations could automatically group more nodes besides 'Music', such as 'Vehicle sounds' or 'Environmental sounds', whose semantic relations could characterise different acoustic environments. 

\subsection{Latent acoustic representation on VAE}

Another contribution of this study is the addition of a VAE for self-supervised regrouping of data in latent space. Hence, the data are not linked to the GPS cell location tags so that the algorithm is able to find disparities in the parameters of the normal distributions.

Results have already been shown in Section \ref{RE}, where the heatmap related to one user location (Fig. \ref{heatmaps}) is able to find a cluster related to metro locations and another cluster for indoor locations, leading to further discriminative latent spaces. 
Also shown in Fig. \ref{scatter} the evolution of data regrouping during epochs for another user, so that a clustering of the data of location number 10 can be observed. It should be noted that the initialization of the t-SNE algorithm is random and this leads to non-deterministic visual results.

The training of a user-specific model focuses on customization and flexibility of the architecture. Training individual models per user provides the flexibility to use different architectures or hyperparameters for each model, optimizing them for the specific characteristics of each user's data. Therefore, each model will be tuned to the particular preferences, patterns and characteristics of a singular user, in line with personalization or customization techniques. However, other options were considered that could also work, e.g., using data from all users as an initialization of the system and then refining it with specific models as more data becomes available.

\section{Conclusions} \label{CON}

In this paper, we try to answer the question of whether it is possible to characterize an unlabeled in-the-wild acoustic scene based on the space-time that it actually covers and a limited portion of user location-labeled scenes. Results presented seem to have achieved a regrouping or clustering of data on the VAE's latent space, based on the space-time that the acoustic scene is related to. 

The main factor that may be influencing the robustness of this analysis is the implementation of the experiment in-the-wild. This leads to a pseudo-labelling of GPS-associated data that inhibits  the assessment of classification metrics, as well as difficulties in the detection of acoustic events that could introduce bias in the results. Furthermore, using both the TF-IDF algorithm (which reduces highly repetitive sounds such as those related to noise) and a $99^{th}$ percentile activation threshold reduces this bias.

In a continuous line, combining NLP techniques as TF-IDF and Node2Vec provides rich information exploitation. As well as reducing the bias with TF-IDF and identifying the relevant acoustic events for an audio with respect to all audios with Node2Vec, which is able to capture the semantic relationships of these. 

Another noteworthy aspect pertains to the comparative analysis of distinct acoustic event detection models. As new models of this type are becoming publicly available it is important to make an assessment of the new contributions. For this \welive dataset it has been shown that YAMNet requires less inference time than PANNs CNN14. 

To conclude, particular emphasis is placed upon the use of the VAE's latent space to characterize acoustic environments as a method of self-supervised learning. This latent space is able to regroup or characterize distinguishable acoustic scenes in space-time for our dataset. 

\section{Future work} \label{FW}

One of the forthcoming steps entails an in-depth analysis of additional pre-trained networks, beyond YAMNet and PANNs CNN14. As mentioned in Section \ref{COMP} there are studies that are working in this direction \cite{PANNs}.

However, since the main goal of EMPATIA-CM project is to detect risk situations in order to prevent GBV, the next step of this branch of the project is to do an affective analysis for acoustic spatio-temporal classification in-the-wild. So far the \welive dataset consists of a limited number of emotion labels, so it has been left out of the study. However, as the \welive data grows --as the research team wants to capture more data in a similar way--, the affective component will be introduced into the project. Furthermore, as this is a collaborative project that also includes the relationship of both physiological and speech signals to emotion analysis, one of the future steps is also to coordinate these modalities for prediction. 

In order to include affective analysis in the study other pipelines could be evaluated, e.g. AudioSet has its own description of each sound event. This description could be treated as a text, and by explaining all acoustic events of one audio there are many NLP transformer-based models that could classify sentiments, such as BERT \cite{BERT}. There are also other NLP models that could classify sentiments based on the AudioSet description as WavCaps \cite{WavCaps} that is assisted by ChatGPT and it is able to filter and transform raw descriptions automatically.

Looking at other recently published studies \cite{Kushwaha}, another line of research that would be useful to address is zero-shot learning. Zero-Shot Learning is an approach in the field of machine learning and artificial intelligence that is used to train models to recognize and understand classes or categories of objects or concepts for which they have not been explicitly trained. In other words, instead of requiring a large set of labeled data for each class to be recognized, zero-shot learning allows a model to generalize and understand new classes without having previously seen them during training \cite{zeroshot}. Given the similarities between this paper and that of \emph{Kushwaha, S. et al.} \cite{Kushwaha}, as both use the latent space at the output of an encoder to characterize acoustic scenes, zero-shot learning would be a promising future line when detecting risk situations in a real spatio-temporal situations, since it is open to new acoustic scenes that may appear in real environments. 

\section*{Acknowledgment}
This work has been partially supported by 
the SAPIENTAE4Bindi Grant PDC2021-121071-I00 funded by MCIN\/AEI\/10.13039/501100011033 and by the European Union ''NextGenerationEU/PRTR'', PID2021-125780NB-I00 funded by AEI, the Spanish Ministry of Science, Innovation and Universities with the POP-FPU grant FPU19/00448. The authors thank all the members of UC3M4Safety for their contribution and support.

\end{document}